# Magnetoelectric control of spin helicity and nonreciprocal charge transport in a multiferroic metal

Daiki Yamaguchi[1,2*], Aki Kitaori[1,3], Naoto Nagaosa[2,4], Yoshinori Tokura[1,2,5*]

[1]*Department of Applied Physics, The University of Tokyo, Tokyo 113-8656, Japan*
[2]*RIKEN Center for Emergent Matter Science (CEMS), Wako 351-0198, Japan*
[3]*Institute of Engineering Innovation, The University of Tokyo, Tokyo 113-0032, Japan*
[4]*RIKEN Fundamental Quantum Science Program, Wako 351-0198, Japan*
[5]*Tokyo College, The University of Tokyo, Tokyo 113-8656, Japan*

DY*: dai.20.yama.011-1@g.ecc.u-tokyo.ac.jp,
AK: kitaori@ap.t.u-tokyo.ac.jp,
NN: nagaosa@riken.jp,
YT*: tokura@riken.jp

* To whom correspondence should be addressed.

**Data and materials availability**

All data needed to evaluate the conclusions in the paper are present in the paper and/or the Supplementary Materials.

**Fundings**

This work was supported in part by JSPS KAKEHNHI (Grant No. 23H05431) and JST PRESTO (Grant No. JPMJPR23Q3).

**Competing interests**

Authors declare that they have no competing interests.

**Abstract**

A multiferroic state with both electric polarization (***P***) and magnetization (***M***) shows the inherently strong ***P-M*** coupling, when ***P*** is induced by cycloidal (Néel-wall like) spin modulation. The sign of ***P*** is determined by clockwise or counterclockwise rotation of spin, termed the spin helicity. Such a multiferroic state is not limited to magnetic insulators but can be broadly observed in conductors. Here, we report the current-induced magnetoelectric control of the multiferroics in a helimagnetic metal YMn$_6$Sn$_6$ and its detection through nonreciprocal resistivity (NRR). The underlying concept is the coupling of the current with the toroidal moment $\boldsymbol{T} \sim \boldsymbol{P} \times \boldsymbol{M} \sim (\hat{\boldsymbol{q}} \times \boldsymbol{\chi}_v) \times \boldsymbol{M}$ as well as with the magneto-chirality $\boldsymbol{\chi}_v \cdot \boldsymbol{M}$, where $\hat{\boldsymbol{q}}$ and $\boldsymbol{\chi}_v$ being the unit modulation wave vector and the vector spin chirality, respectively. We furthermore observe an enhancement of NRR by the spin-cluster scattering via $\boldsymbol{\chi}_v$ and its fluctuation. These findings may pave a way to exploration of multiferroic conductors and application of the spin-helicity degree of freedom as a state-variable.

**Introduction**

The chirality is one of the fundamental notions in a broad field of science. In condensed matter science, chirality in magnetism has been attracting enormous interest in terms of multiferroics, skyrmions, nonreciprocal responses and their application to spintronic devices. For example, the chirality can be found in helical magnetic structures, such as proper screw or Bloch-wall like (Fig. 1E) and cycloidal or Néel-wall like (Fig. 1G (plan view)) spin states. Here, the term *spin helicity* refers to the rotation sense of spins in proceeding along the modulation wave vector ***q***. The spin helicity in helimagnetic structures is reversed by a mirror operation.

Multiferroics are the materials, in which multiple ferroic orders, such as ferroelectricity, ferroelasticity and ferromagnetism, coexist (*1*). In particular, the coupling of ferroelectricity and ferromagnetism/antiferromagnetism has been intensively studied in the last few decades. In insulating multiferroics, the mutual control of magnetization (***M***) and electric polarization (***P***) by electric and magnetic fields, respectively, is realized via the magnetoelectric (ME) effect. Multiferroics of spin origin can emerge in some helical magnetic structures with spin-helicity degree of freedom. A microscopic mechanism to describe the generation of ***P*** from noncolinear spin modulation has been proposed, termed the spin current model (*2*) or inverse Dzyaloshinskii-Moriya model (*3, 4*). In this model, the emerging ***P*** is expressed as

$$\boldsymbol{P} = A\sum \boldsymbol{e}_{ij} \times \chi_v^{ij} = A\sum \boldsymbol{e}_{ij} \times (\boldsymbol{S}_i \times \boldsymbol{S}_j), \qquad (1)$$

where *A* is a coupling constant which reflects the spin exchange interaction and spin-orbit coupling (SOC), and ***e***$_{ij}$ is the unit vector connecting neighbouring-site spins ***S***$_i$ and ***S***$_j$. $\chi_v^{ij} = \boldsymbol{S}_i \times \boldsymbol{S}_j$ is the vector spin chirality. This formula ensures that the canting modulation of spins produces the local ***P***. The local ***P*** remains macroscopically finite in some helimagnetic structures such as the cycloidal spin structure (Fig. 1G (plan view)). In the cycloidal spin structure, the neighbouring-site spins are stabilized at an angle in the spin rotation plane, which is parallel to the modulation direction (***q*** vector). (In the case of constant-pitch helical spins, the vector spin chirality $\chi_v$ takes a value independent of site pair *ij*.) This mechanism has been experimentally confirmed in various magnetic insulators (*1*). The direction of ***P*** depends on the spin helicity of cycloid, *i.e.*, the direction of cycloidal spin rotation. It is demonstrated that the spin-helicity control by the electric and magnetic fields switches the direction of ***P*** in multiferroic (Mott) insulators (*5, 6*). One of the motivations of the present work is to test the generation of such a ferroelectric polarization in a metal with the cycloidal spin state and its ME control in analogy to the multiferroic insulators. However, the electric-field control of ***P*** (or spin helicity) appears impossible in a metal because of strong dielectric screening, instead we try to exploit the toroidal moment which is an order parameter characteristic of multiferroics and can couple with the flowing current in metallic multiferroics.

Helimagnets have recently been attracting attentions in terms of the spintronic application for a novel magnetic memory or computing device (*7-9*). For example, spin-helicity degree of freedom has a potential as a state-variable in helimagnets. To this end, the spin helicity should be controlled and detected electrically in conducting helimagnets. This is recently successfully achieved in a centrosymmetric itinerant proper-screw helimagnet MnP (*8*), followed by the demonstration of the spin-helicity control for proper-screw MnAu$_2$ thin film at room temperature (*9*). In these cases, the electrical magnetochiral effect (eMChE) (*10-12*) was used to control and detect the spin helicity. The control parameter for eMChE is $\chi_v \cdot \boldsymbol{M}$ (*13*), where ***M*** is the spontaneous or magnetic-field induced magnetization reflecting the conduction electron's spin polarization. The quantity $\boldsymbol{\chi}_v \cdot \boldsymbol{M}$, hereafter referred to as *magneto-chirality*, in the ordered longitudinal (Fig. 1F) or transverse (Fig. 1G) conical structures can cause the electronic-band asymmetry, even without relativistic SOC, which directly produces the nonreciprocal charge transport depending on the directions of current and $\boldsymbol{M}$ (*12*). To control the spin helicity in the proper screw structure (Fig. 1E), for example, a dc current and a magnetic field causing $\boldsymbol{M}$ are simultaneously applied either parallel or antiparallel to stabilize the single-spin-helicity domains (magnetochiral (MCh)-poling). The nonreciprocal resistivity (NRR) due to the eMChE is used to detect the single or dominant spin helicity. Note here again that the eMChE due to the action of $\boldsymbol{\chi}_v \cdot \boldsymbol{M}$ (*13*) does not require the relativistic SOC of the materials system.

In this study, in addition to such an eMChE, we seek for a new effect for control and detection of the spin helicity in a multiferroic metal in terms of ME effect based on the SOC. The target magnetic structure is the transverse conical (TC) state (Fig. 1G), in which $\boldsymbol{P}$ from cycloidal spin modulation and the net magnetization ($\boldsymbol{M} \perp \boldsymbol{P}, \boldsymbol{q}$) coexist, *i.e.*, multiferroic (*14*). For the switching of spin helicity or $\boldsymbol{P}$ direction in the TC state, we exploit the toroidal moment $\boldsymbol{T}$ which can couple with the current $j$; the toroidal moment is defined as $\boldsymbol{T} = \frac{1}{2}\sum_i \boldsymbol{r}_i \times \boldsymbol{S}_i$, where $r_i$ and $S_i$ are position and spin moment vectors at site $i$ (*15*), and hence in the multiferroic state is approximated as $\boldsymbol{T} \sim \boldsymbol{P} \times \boldsymbol{M}$ in the magnetically-ordered electrically-polar state. Combined with the expression of $\boldsymbol{P}$ in Eq. (1), $\boldsymbol{T}$ in the general conical structure can be expressed as $\boldsymbol{T} \sim (\hat{\boldsymbol{q}} \times \boldsymbol{\chi}_v) \times \boldsymbol{M}$ with $\hat{\boldsymbol{q}}$ being the unit vector along the spin modulation $\boldsymbol{q}$ vector. This should show a distinct $\boldsymbol{M}$-directional dependence from the case of the magneto-chirality $\boldsymbol{\chi}_v \cdot \boldsymbol{M}$. Here, we report that the multiferroic state (TC state) in a metallic magnet YMn$_6$Sn$_6$ can be controlled electrically via such an ME effect of $\boldsymbol{T}$ as well. Moreover, the controlled multiferroic state is detected through a novel NRR arising from $\boldsymbol{T}$ as well as from the eMChE, accompanied by the large NRR from the electron scattering from the spin cluster with fluctuating vector spin chirality near magnetic phase boundaries.

**Results**

**Helimagnetic orders of YMn$_6$Sn$_6$.** The single crystal of YMn$_6$Sn$_6$ has a centrosymmetric hexagonal structure (space group *P*6/*mmm*, No. 191) with *a* = *b* = 5.54 Å and *c* = 9.01 Å (Fig. 1A). The frustrated exchange interactions between 3*d* moments of Mn kagome lattices produce rich magnetic phase diagrams with some helical magnetic structures with the ***q*** vector along the *c*-axis (Figs. 1, C and D). There are interlayer interactions, ferromagnetic and antiferromagnetic, between the two kinds of nearest-neighbor Mn-plane spins, which stabilizes an up-up-down-down like double-antiferromagnetic structure. In addition, the second-nearest-neighbor ferromagnetic interaction between the Mn-plane spins along the *c*-axis gives a frustration to this commensurate order and stabilizes helical magnetic structures. Thus, the helix in this material has a distorted or a double spiral structure, with two different spin rotation angles. The helical pitch is short in this material, around 3 nm. The proper screw (PS) magnetic structure (Fig. 1E) stabilizes in zero field over a wide range of temperature between the lowest temperature (< 2 K) and $T_N$ = 330 K (*16-19*). Upon increasing the magnetic field applied along the *c*-axis, PS becomes the longitudinal conical (LC) structure (Fig. 1F) because of tilt of spins along the magnetic field. The LC finally transforms into the forcedly ferromagnetic (FF) phase (Fig. 1I) above a threshold magnetic field. On the other hand, with applying the magnetic field along the *a*-axis, spin flop transition from the PS to the transverse conical (TC) structure (Fig. 1G) occurs at around *B* = 2.5 T, where the spin rotation plane flops its direction by 90 degrees, *i.e.*, from the *ab*-plane to the *a*\**c*-plane; here *a*\* is set perpendicular to the *a*-axis within the *ab*-plane. These helical magnetic structures, PS, LC and TC states, show the spin-helicity degree of freedom in the originally centrosymmetric chemical lattice. Aside from these helical magnetic structures, there is a fan like (FL) phase (Fig. 1H) above 7 T below 150 K. Just below the phase boundary at $T_N$, a mixture of antiferromagnetic order (AF) and PS, LC or TC structure is observed (*20-24*). The temperature dependence of the longitudinal resistivity along the *c*-axis is measured in 0 T and 5 T (***B***//*a*) (Fig. 1B), showing a low residual resistivity (< 10 μΩcm) and good metallicity. The kinks at 300 K and 270 K in the 5 T curve correspond to the FF-AF and AF-TC phase transitions.

Figures 1E-1G further elaborate the magnetic structure of PS, LC and TC phases along with the definition of spin helicity. (Hereafter, the actual double helix structure is approximated by the single helix presentation, which does not essentially change the following discussion of the results.) In the PS and LC structures, the rotation plane of the magnetic moment is perpendicular to the *q* vector (//*c*). The spin helicity is defined as right-handed or left-handed, depending on the direction of spin rotation viewed along the *q* vector. On the other hand, the external magnetic field (> 2.5 T) applied perpendicular to the *q* vector stabilizes the TC, in which the rotation plane is parallel to the *q* vector. The spin helicity is defined as clockwise (CW) or counterclockwise (CCW), depending on the direction of the spin rotation of the cycloidal component viewed along the cone-axis of TC in proceeding along the *q* vector (//*c*). Without any procedure to create the single-spin-helicity domain using current application, PS and TC structures in YMn$_6$Sn$_6$ take the multi-domains of spin helicity with the nearly equal population due to the energetical degeneracy of the two spin-helicity states, as evidenced by almost indiscernible NRR behaviour, *vide infra*.

One distinct feature in TC phase is the existence of finite ***P*** as described by Eq. (1), *i.e.*, multiferroic in nature. Note that the direction of ***P*** is determined by the spin helicity of the cycloidal component of TC, *i.e.*, CW or CCW. Moreover, TC has the magnetization component (***M***) along the cone-axis, which is perpendicular to both ***P*** and ***q*** vectors. The toroidal moment $\boldsymbol{T} \sim \boldsymbol{P} \times \boldsymbol{M} \sim (\hat{\boldsymbol{q}} \times \chi_v) \times \boldsymbol{M}$ is along the *q* vector (//*c*), and the sign of ***T*** is determined by the sign (CW or CCW) of spin helicity, or equivalently of ***P***, when ***M*** is fixed (Fig. 1G). ***T*** can be viewed as a built-in vector potential $\boldsymbol{A}_{\text{eff}}$ under the SOC, since the effective SOC term in Hamiltonian is described as

$$\lambda \boldsymbol{L} \cdot \boldsymbol{S} = \lambda (\boldsymbol{r} \times \boldsymbol{p}) \cdot \boldsymbol{S} = -\lambda (\boldsymbol{r} \times \boldsymbol{S}) \cdot \boldsymbol{p} = e \boldsymbol{A}_{\text{eff}} \cdot \boldsymbol{p}, \qquad (2)$$

where $\lambda$, ***L***, *e* and ***p*** are the spin-orbit interaction, the angular momentum, the electron's elementary charge and momentum, respectively (*25*). Note that unlike the TC phase the PS or LC phase configurations do not have finite ***P*** or ***T*** according to Eq. (1).

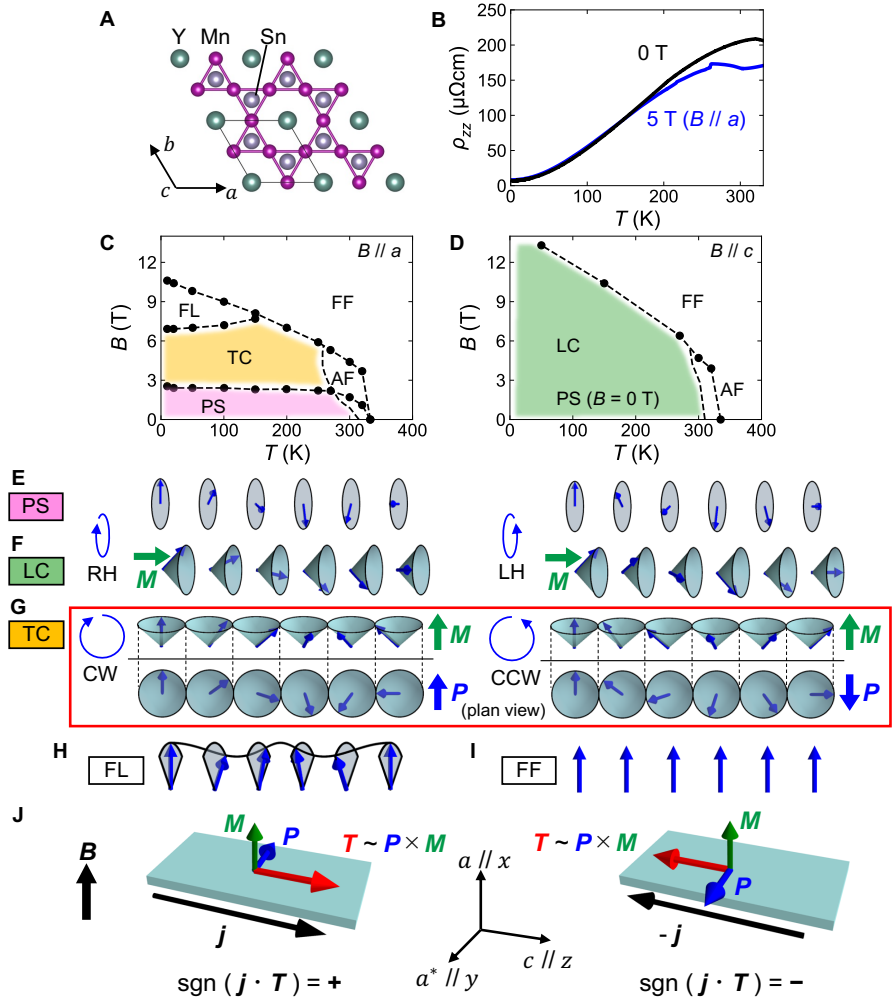

**Fig. 1 Crystal and magnetic structures of YMn$_6$Sn$_6$ with definition of spin helicity.** (**A**) Schematic crystal structure of YMn$_6$Sn$_6$ viewed along the *c*-axis, drawn by using VESTA (*34*) (**B**) Temperature dependence of longitudinal resistivity along *c*-axis in 0 T (black) and 5 T (***B***//*a*) (blue). Magnetic phase diagram of YMn$_6$Sn$_6$ with (**C**) ***B***//*a* and (**D**) ***B***//*c*. PS, LC, TC, AF, FL and FF stand for proper screw, longitudinal conical, transverse conical, antiferromagnetic, fan like and forcedly ferromagnetic structures, respectively. Schematics of (**E**) PS and (**F**) LC structures. PS and LC of opposite spin helicities, *i.e.*, right-handed (RH) and left-handed (LH) screws are shown. (**G**) Schematic of TC structure. Upper and lower show the side and plan views of TC structure, respectively. Clockwise (CW) and counterclockwise (CCW) spin helicities are shown. Schematics of (**H**) FL and (**I**) FF structures, where spin helicity is not defined. (**J**) Direction of toroidal moment ***T*** (red arrows) and electric current ***j*** (black arrows) in magnetoelectric (ME)-poling process for CW and CCW modulations, respectively. The light-blue plate represents the sample. TC phase is stabilized by the application of external magnetic field ***B***//*a*. The direction of spontaneous magnetization ***M*** and electric polarization ***P*** are indicated by green and blue arrows in (F, G).

**Magnetoelectric poling of spin helicity.** In this study, we exploit the toroidal moment along the *q* vector ($T_z$) in the TC phase to control and detect the spin helicity. ***T*** as a built-in vector potential couples with the electron's momentum ***p*** or electric current (Eq. (2)). This means that the electronic-band asymmetry with respect to the $k_z$ direction should arise in the TC phase, in a manner depending on the sign of the spin helicity or ***P***, and accordingly shows the nonreciprocal electron transport as determined by the sign of $T_z$, as shown in Fig. 1J. Conversely, the electric current can align $T_z$ and hence ***P*** and spin helicity in the case of a multi-helicity domain state. This ME coupling between the current and the spin helicity is analogous to the action of magneto-chirality $\chi_v \cdot \boldsymbol{M}$, which are both expected to cause the electronic-band asymmetry. However, the former (latter) is relevant (irrelevant) to the relativistic SOC, and the ***M***-directional dependence is distinct from each other.

The current induced poling of ferrotoroidic domain was previously demonstrated for an antiferromagnet with finite toroidal moment (*26*). In a similar manner, it is expected that $T_z$ in the present multiferroics with the ordered TC structure can also be switched by electric current density (*j*). We adopted a new procedure to control the spin helicity with such a $T_z$-*j* coupling mechanism as the working hypothesis, as termed *magnetoelectric (ME)-poling*. Note here that 'ME' refers to the magnetoelectric coupling, in which the toroidal moment is an essential order parameter, and we applied electric current instead of electric field, in contrast to the conventional poling procedure for insulating multiferroics. As shown in Fig. 2A, at first, we applied the magnetic field along the *a*-axis (perpendicular to the ***q***//*c*) up to +7 T at 250 K, and the high dc current density $j_{pol}$ along the *c*-axis. Here, $j_{pol}$ in the opposite direction is anticipated to realize the opposite helicity of TC phase in the end. We *tentatively* define the spin helicity of TC as CW and CCW after the ME-poling with positive and negative $j_{pol}$, respectively. (We can know whether the spin helicity is reversed or not between CW and CCW, but cannot distinguish which is CW or CCW from the present nonlinear transport experiment alone, since the sign of $T_z$ and/or $\chi_v \cdot \boldsymbol{M}$ of CW or CCW depends on the material parameters, such as the coefficient *A* in Eq. (1).) Then, magnetic field was slowly removed at a reduction rate of 1.2 mTs$^{-1}$ from 7 T to 5 T while the current of $j_{\text{pol}}$ was kept flowing, enabling to traverse the phase boundary from the FF phase into the TC phase (Fig. 2A).

The spin helicity of the TC state is detected by the nonreciprocal charge transport arising from $T_z$ and/or $\chi_v \cdot \mathbf{M}$. The sign of the NRR changes depending on the sign of $T_z$ and/or $\chi_v \cdot \mathbf{M}$. Here the NRR is defined as an additional term ($\rho_{\text{ch}} j$) in the resistivity, up to the first order of $j$,

$$\rho(j) = \rho_0 + \rho_{\text{ch}} j, \tag{3}$$

where $\rho_0$ and $\rho_{\text{ch}}$ are linear resistivity and constant related to NRR. The NRR can be measured as a second harmonic ($2\omega$) resistivity in the lock-in measurement while applying the ac current ($j = j_{\text{ac}} \sin(\omega t)$) with frequency ($f = \omega/2\pi$),

$$-\rho_{\text{ch}} j_{\text{ac}} \cos(2\omega t) = \rho^{2f} \cos(2\omega t). \tag{4}$$

The NRR refers to $\rho^{2f}$. The NRR signal is finite when $T_z$ and/or $\chi_v \cdot \mathbf{M}$ is finite. Therefore, the single spin helicity realized by the ME-poling can be detected by the finite NRR in TC phase which becomes saturated as increasing $j_{\text{pol}}$. Hereafter, we call this method *NRR sensing*. As illustrated in Fig. 2A, the NRR sensing was performed at 50 K after the ME-poling at 250 K. Once the ME-poling was completed at 250 K, the system was cooled down to 50 K in 5 T. Then, the field was raised up to 6 T, which is just below the phase boundary between TC and FL (see Fig. 1C). The NRR was measured while the field was swept between 6 T and 0 T. During the field sweep, ac electric current with $j_{\text{ac}} = 5 \times 10^8$ Am$^{-2}$ and $f = 1$ kHz was applied along the $c$-axis, as a probe excitation to detect the NRR.

Figure 2B shows the representative result of the NRR sensing after ME-poling with $j_{\text{pol}} = \pm 5 \times 10^8$ Am$^{-2}$. To correct some background extrinsic contributions to nonreciprocal signals, we defined the extrinsic background as an average of CW ($+j_{\text{pol}}$) and CCW ($-j_{\text{pol}}$) signals, and then we derived the NRR for $\pm j_{\text{pol}}$ as deviations from this background level; see the section S1 in Supplementary Materials (SM) for the detailed procedure of data analysis and its justification. (Thus, the CW and CCW NRRs shown in the figures show up as a plus-minus symmetric curves by definition.) At the beginning of the field sweep from 6 T, the system is in the TC phase, and the NRR is finite at around +0.2 and -0.2 nΩcm for CW and CCW states, respectively. The magnitude of the signal keeps its value until the system undergoes the phase transition at 2.5 T to the PS phase.

The NRR value is confirmed to be linear to the probe ac current density $j_{ac}$ as expected from Eq. (4) (see S2 in SM). After the phase transition into the PS phase is completed with the decrease of magnetic field, the NRR becomes nearly zero. This is consistent with the expectation that $T_z$ and $\chi_v \cdot M$ are finite in the TC phase but both are zero in the PS phase. (Note that $\chi_v \cdot M$ is also zero in the present case since $\chi_v \perp M$ in the PS phase with $M//x$.) In the course of the TC-to-PS transition, however, a large sharp peak structure of NRR is observed at the phase boundary immediately below 2.5 T, which may arise from the effect of asymmetric charge-carrier scattering on the residual or fluctuating chiral spin clusters (*13*) in the PS phase close to the boundary with the TC phase. This NRR response is distinct in the mechanism from the NRR in the long-range ordered TC phase and will be elaborated in the following section.

Figure 2C shows the $j_{pol}$ dependence of the magnitude of NRR. With small $j_{pol}$ = $1\times10^7$ Am$^{-2}$, the NRR in TC phase is not discerned (see Fig. S1D in SM), showing that the multi-domains of spin helicity with the nearly equal population of the CW and CCW states. Upon increasing the $j_{pol}$, the magnitude of NRR increases and then saturates at around $j_{pol}^{th}$ = $7\times10^7$ Am$^{-2}$. The saturation of the NRR indicates that the single-helicity domain is realized above the $j_{pol}^{th}$. Thus, the ME-poling and the NRR sensing using the toroidal moment $T_z$ and/or the magneto-chirality $\chi_v \cdot M$ is shown to be applicable to control and detect the spin helicity in the multiferroic TC metal state. As for the spin-helicity control of the PS or LC phase, the MCh-poling, which can be done via poling under the current flow and the external magnetic field both along the *q* direction (*c*-axis), is effective as shown in previous studies (*8*, *9*), while the $T_z$-based ME-poling is not applicable for the LC phase. In fact, we could achieve the similar MCh-poling and NRR sensing of the PS/LC phase, as described in S3 in SM. In Fig. 2C is also shown the $j_{pol}$ dependence of the magnitude of NRR for the MCh-poling; the $j_{pol}^{th}$ in this case is around $1\times10^8$ Am$^{-2}$, appreciably larger than the case of the ME-poling, while the saturated NRR value is larger, implying that the microscopic poling mechanism is more or less different between the ME- and MCh- poling as anticipated.

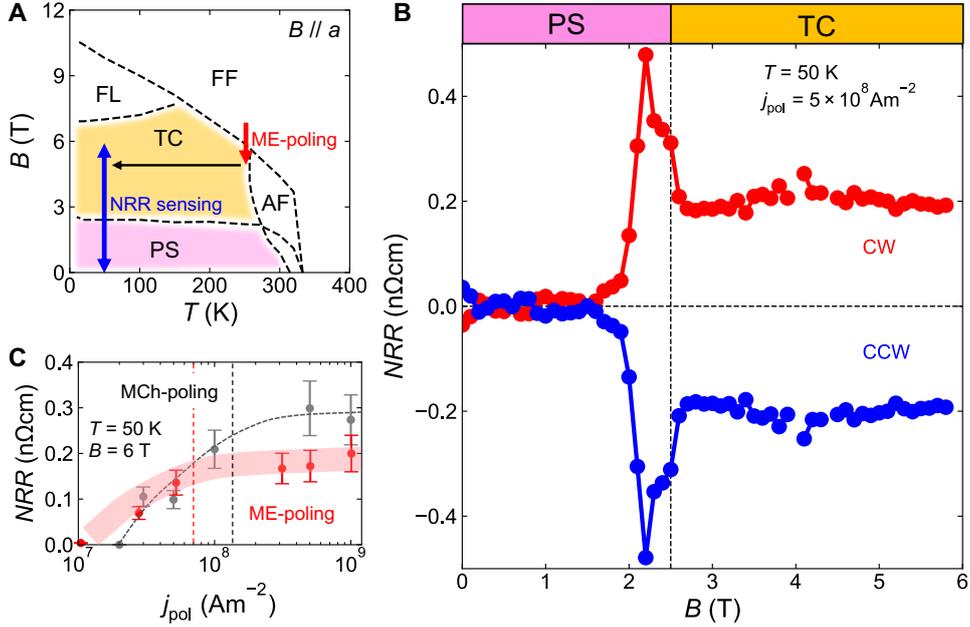

**Fig. 2 Control and detection of multiferroicity in YMn$_6$Sn$_6$.** (**A**) Magnetoelectric (ME)-poling (red arrow) and nonreciprocal resistivity (NRR) sensing (blue arrow) procedures are indicated in the magnetic phase diagram. (**B**) Magnetic field dependence of the NRR measured at 50 K with $j_{ac}$ = 5×10$^8$ Am$^{-2}$ after ME-poling with $j_{pol}$ = +5×10$^8$ Am$^{-2}$ (red, CW) and $j_{pol}$ = -5×10$^8$ Am$^{-2}$ (blue, CCW). (**C**) Poling current density $j_{pol}$ dependence of NRR for ME-poling (red) and mangetochiral (MCh)-poling (gray). The red and black vertical lines approximately indicate the threshold $j_{pol}^{th}$ for ME (for TC) and MCh (for LC) configurations, respectively. The red fat line and the gray dashed curve are the guides to the eyes.

**Angular dependence of NRR.** The NRR sensing is performed in various magnetic field sweep directions after the ME-poling, to establish the relative contribution of the above two distinct mechanisms, i.e., based on $T_z$ and $\chi_v \cdot M$, to the NRR in TC phase. The magnetic field is swept from 0 T to 6 T at various deviation-angles $\theta$ from the $a$-axis within the $ac$-plane, as shown in the left panel of Fig. 3A: $\theta$ = 0° corresponds to the limit where $T_z$ takes the maximum, and $\theta$ = 90° to the limit where $T_z$ is zero. The NRR of the $T_z$ origin (*27*) is termed hereafter *electrical toroidal-chiral effect* (eTChE). At $\theta$ = 0°, the NRR is from both finite $T_z$ and $\chi_v \cdot M$ in the TC phase. On the other hand, at $\theta$ = 90°, where $T_z$ = 0, the NRR is arising solely from the eMChE in the LC phase. At angles in between, the NRR may appear from both of these effects. With increasing $\theta$ from 0°, the NRR in the low-field region below 2 T gradually increases toward the peak structure at around 2.5 T. The peak structure at $\theta = 15°$ is not as sharp as compared to that at $\theta = 0°$ where the PS-TC spin flop transition is well-defined. Upon increasing the tilt angle, the peak structure becomes even more dull toward $\theta = 30°$ and the NRR increases almost monotonically at angles $\theta \gtrsim 45°$.

The observed NRR magnitudes at 50 K and at selected values of magnetic field $B = 1.5$ T, 3 T and 6 T are plotted in Fig. 3B as a function of the angle $\theta$ of $\boldsymbol{B}$ from the $a$-axis. The respective behaviors are predicted by a simple calculation (see S4 in SM) of the angular dependence of $\boldsymbol{\chi}_v \cdot \boldsymbol{M}$ and $T_z \sim [(\hat{\boldsymbol{q}} \times \boldsymbol{\chi}_v) \times \boldsymbol{M}]_z$, as shown in Fig. 3C. Among these, $\boldsymbol{\chi}_v \cdot \boldsymbol{M}$ does not have strong angular dependence at 3 T and 6 T in the course of TC-to-LC phase change, where the spin cone-axis continuously rotates in the ac plane as schematically shown in the left panel of Fig. 3A. However, $\boldsymbol{\chi}_v \cdot \boldsymbol{M}$ at 1.5 T is zero in the genuine PS phase at $\theta = 0°$ (see Fig. 1C) and gradually increases with $\theta$ toward the LC state (see Fig. 1D). This behavior depends on the critical field strength for the spin flop transition at general $\theta$ (2.5 T at $\theta = 0°$) which is not easily estimated. Therefore, the continuous rotation of the cone-axis is tentatively assumed also in the calculation at 1.5 T between $\theta = 15°$ and 90°. On the contrary, $T_z$ is strongly suppressed toward $\theta = 90°$ from its maximum value at $\theta = 0°$ (Fig. 3C). $T_z$ appears almost doubled with $B$ changing from 3 T to 6 T, while $T_z$ in 1.5 T is zero in the PS phase ($\theta = 0°$). For angles between $\theta = 15°$ and 90°, $T_z$ is again calculated assuming the continuous cone-axis rotation, and vanishingly small values are obtained due to the large cone-opening angle at 1.5 T. The angular dependence of the NRR in 3 T and 6 T (all the cases where the contribution from the spin-cluster scattering is small enough) are fit by the superposition of these two amounts ($NRR = a\boldsymbol{\chi}_v \cdot \boldsymbol{M} + bT_z$) as shown in Fig. 3B. The coefficients are found to be $a = 0.33$ and $b = -0.16$, indicating the opposite-sign contribution from $\boldsymbol{\chi}_v \cdot \boldsymbol{M}$ and $T_z$ with the comparable absolute magnitude, *i.e.*, 2:1. In this way, the eMChE and eTChE can be separably estimated. In particular, the NRR in the TC phase of YMn$_6$Sn$_6$ partly arises from a newly assigned mechanism, namely eTChE based on SOC, which is different from the conventional eMChE observed in the LC phase. Note that the sign of the resultant NRR in the present case is more dominated by eMChE than by eTChE, however the situation will depend on the actual materials parameters for the TC phase, in particular on the magnitude of the relativistic SOC favoring the eTChE.

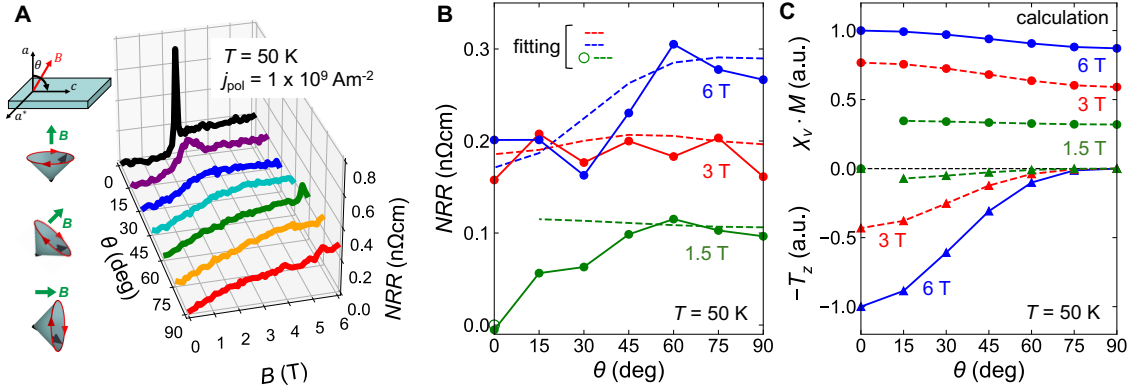

**Fig. 3 Angular dependence of nonreciprocal resistivity (NRR).** (**A**) Magnetic field dependence of NRR under various magnetic field directions within the *ac*-plane. The NRR sensing is performed at 50 K in various magnetic field directions with $j_{ac} = 5×10^8$ **Am$^{-2}$** after ME-poling with $j_{pol} = 1×10^9$ **Am$^{-2}$**. Schematics show the definition of the magnetic field rotation angle $\theta$, and the cone axis directions in TC phase at $\theta = 0°$ (***B***//*a*), 45° and 90° (***B***//*c*). (**B**) $\theta$ dependence of NRR in 6 T (blue), 3 T (red) and 1.5 T (green) at 50 K. Blue, red and green dashed lines are the fitting by the superposition of the calculated magneto-chirality $\chi_v \cdot M = (S_i \times S_j) \cdot M$ and the toroidal moment along the ***q*** vector $T_z \sim [(\hat{q} \times \chi_v) \times M]_z$ in (C). (See the main text for the detail.) The fitting for 1.5 T data is difficult due to the PS-LC spin flop transition at low $\theta$ region and hence tentatively shown between $\theta = 15°$ and 90°. Fitting for $\theta = 0°$ is shown separately with the green open circle. (**C**) Calculated $\theta$ dependence of $\chi_v \cdot M$ and $T_z$ at 50 K. Blue, red and green correspond to the values in 6 T, 3 T and 1.5 T, respectively.

**Enhancement of NRR by spin-cluster scattering at phase boundaries.** The observed NRR phenomena in the helimagnetically ordered phases of PS and TC, the NRR appears to arise from the electronic-band dispersion asymmetry between $+k_z$ and $-k_z$ due to the presence of $T_z$ and/or $\chi_v \cdot M$; here $z$ is the axis along helical ***q***-vector (//*c*). Apart from these NRR signals in the ordered phases, the conspicuous NRR scattering peak structure is observed at the PS-TC phase boundary, as shown in Fig. 2B. We assign this conspicuous NRR peak not to the electronic-band dispersion asymmetry effect but to the nonreciprocal scattering from the spin cluster with the vector spin chirality $\chi_v = S_i \times S_j$ (*13*), embedded in the paramagnetic or non-NRR ordered phase (*e.g.*, the PS phase). Such a scattering-effect NRR signal was typically observed for a chiral magnet MnSi near above the helical magnetic transition temperature under magnetic field (*28*).

Figure 4A shows the NRR response in a whole magnetic field region of 0-14 T for $\boldsymbol{B}//a$ ($\theta = 0°$). It is worth noting that the spin-cluster scattering seems to occur at the respective phase boundaries, not only at PS-TC but also at FL-FF boundaries, and that the memory effect of spin helicity subsists even in the FF phase, as follows: The NRR is measured at 50 K after ME-poling with $j_{\text{pol}} = \pm 1 \times 10^9$ Am$^{-2}$. The magnetic field ($\theta = 0°$) is first increased from 0 T to 14 T, traversing the magnetic phases in the order of PS, TC, FL and FF. Then, the field is decreased from 14 T to 0 T. In the upward sweep, the NRR becomes finite after the sharp peak structure at 2.5 T until the TC-FL phase boundary at 9 T. Then, in the FL phase, where the spin helicity is not well-defined, the NRR takes nearly zero value. At the FL-FF phase boundary, however, the NRR again shows a spin-helicity dependent, *i.e.*, poling-history dependent, peak. The FL and FF phases in the well-ordered state should not show NRR. Near the phase boundary, however, the domain walls between FF and FL domains are anticipated to form with the non-collinear spins, *i.e.*, vector spin chirality, and perhaps work as scattering centers of NRR. Interestingly, the memory of the spin helicity appears to be kept even in magnetic phases where the spin helicity appears not to be defined, *i.e.*, FL and FF phases. The NRR is again approaching zero in the FF phase with further increase of magnetic field. Surprisingly, the magnetic field dependence of the NRR in the downward sweep follows the nearly same curve as in the upward sweep. This may be because the spin helicity is memorized possibly by the thermal fluctuation of spins (*29*) or the residual helical spin clusters even in the FL or the FF phases. A similar NRR peak of scattering-origin is also observed for the phase boundary between LC and FF in the case of $\boldsymbol{B}//c$ ($\theta = 90°$), as shown in Fig. 4B.

The sharp enhancement of NRR at the PS-TC phase boundary is further elaborated. Figure 4C shows the magnetic field dependence of the NRR measured at various temperatures after the ME-poling. The temperature dependence of the peak value of the NRR (scattering term) at PS-TC phase boundary at $B = 2.5$ T and the magnitude of the NRR (electronic-band asymmetry term) in the long-range ordered TC phase averaged over 3-6 T are plotted in Fig. 4D. The scattering peak value is conspicuously increased with the temperature, and takes the maximum at 210 K. The large temperature effect of the NRR of scattering origin is due to the thermal spin fluctuation which contributes to the enhanced averaged value of the vector spin chirality $<\chi_v>$ (*13, 28*). On the other hand, the NRR from eTChE and eMChE of the asymmetric band-dispersion origin in the long-range ordered TC phase is almost constant at low temperatures below 150 K, while slightly enhanced toward high temperature.

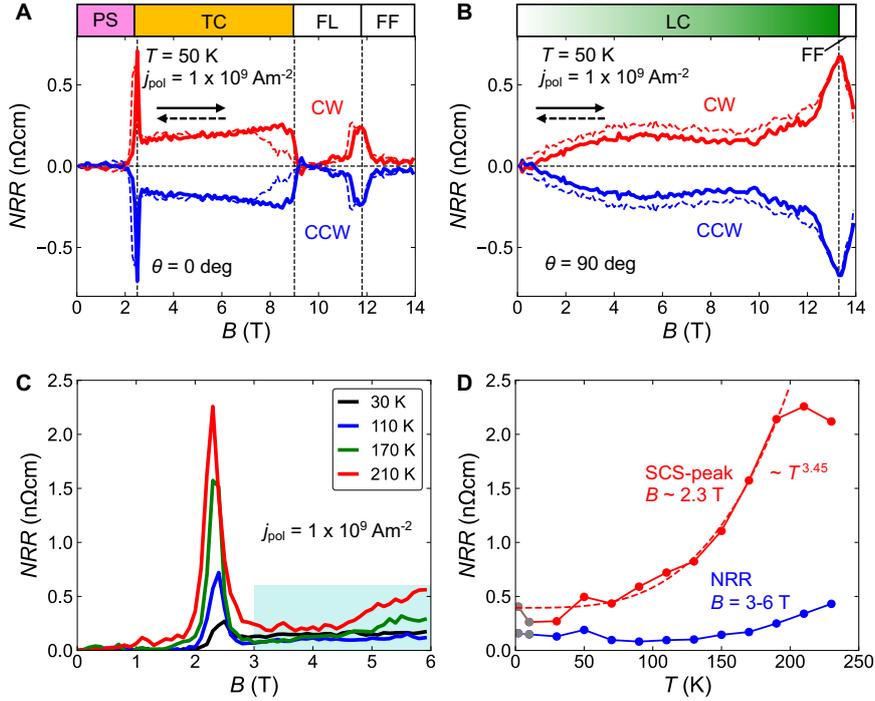

**Fig. 4 Enhancement of nonreciprocal resistivity (NRR) at phase boundaries.** Magnetic field dependence of NRR up to 14 T at (**A**) $\theta = 0°$ and (**B**) $\theta = 90°$ measured at 50 K with $j_{ac} = 5\times10^8$ **Am$^{-2}$** after ME-poling with $j_{pol} = +1\times10^9$ **Am$^{-2}$** (red, CW) and $j_{pol} = -1\times10^9$ **Am$^{-2}$** (blue, CCW), respectively. Solid and dashed lines indicate the upward and downward magnetic field sweep directions, respectively. The upward sweeps are performed first just after the ME-poling procedures. (**C**) Magnetic field ($B//a$) dependence of NRR measured at various temperatures with $j_{ac} = 5\times10^8$ **Am$^{-2}$** after ME-poling with $j_{pol} = +1\times10^9$ **Am$^{-2}$**. (**D**) Temperature dependence of NRR value at peak of spin-cluster scattering (SCS) at PS-TC phase boundary (red) and NRR value of electronic-band asymmetry origin in TC phase (blue). The red dashed line shows the fitting with $NRR = a + bT^n$ ($n=3.45$). The data points shown in gray are not enough reliable due to the contribution from the Joule heating in the sample.

## Conclusion

We have demonstrated the spin-helicity control and detection in a high temperature helimagnet YMn$_6$Sn$_6$ in its multiferroic transverse conical (TC) phase in terms of the electrical toroidal-chiral effect (eTChE) based on the coupling between toroidal moment and electric current, in addition to the conventional electrical magnetochiral effect (eMChE). We utilized the NRR to detect the controlled spin helicity. In addition, the thermally-enhanced large NRR is observed at the magnetic phase boundaries between the proper screw (PS) and TC phases as well as between longitudinal conical (LC) and forced ferromagnetic (FF) phases, presumably due to the nonreciprocal scattering effect from the spin cluster with vector spin chirality. The spin-helicity control achieved in the TC phase means the control of the electric polarization constituting to the toroidal moment in the metal as induced by the cycloidal spin order. Having shown that the spin-helicity control in a metal is possible in the magnetoelectric configuration, the scope of candidate materials for spintronics based on the spin-helicity degree of freedom is now expanded. For example, there is a possibility that one can control the spin helicity of nanometric skyrmions (*30*) irrespective of Néel-wall type (cycloid) or Bloch-wall type (screw) even in centrosymmetric metallic materials (*31*) in terms of the ME- or MCh-poling, as shown here, without changing the chemical composition of a material (*32*) or applying a strain (*33*). In a broader perspective, the exploration of multiferroic conductors with control of emergent electric polarization and toroidal moment, would pave a way to new spintronic materials.

**Experimental Section/Methods**

**Single crystal growth.** Single crystals of $YMn_6Sn_6$ were synthesized by a Sn flux method (*17*). A mixture of ingredient elements with a molar ratio of Y:Mn:Sn = 1:6:30 was put in an aluminium oxide crucible and then sealed in an evacuated quartz tube. The tube was heated up to 1050 °C, subsequently cooled down slowly to 600 °C in an electric furnace. When the furnace reached 600 °C, excess flux was centrifuged. The obtained single crystals with the well-developed facet structures were soaked in hydrochloric solution to remove remaining flux. The single crystallinity was confirmed by Laue x-ray diffraction. No impurity phase was detected by powder x-ray diffraction.

**Focused ion beam (FIB) device fabrication.** A thin plate ($10\times20\times1$ $\mu m^3$) of the single crystal was cut out by using the focused ion beam (FIB) fabrication technique (NB-5000, Hitachi). The plate was mounted on a silicon substrate with patterned Au/Ti-bilayer electrodes. The plate was electrically connected to the electrodes by FIB-assisted tungsten deposition. The plate was finally fabricated into a Hall-bar shape.

**Transport measurements.** The control and detection of spin helicity were performed in the Physical Property Measurement System (PPMS, Quantum Design). dc electric current was applied to the FIB device through a current source (K6221, Keithley) during the ME-poling procedure. The nonreciprocal resistivity (NRR) was measured by lock-in amplifiers (LI5660, NF corporation). ac current of 1 kHz was applied to detect the second harmonic resistivity. The voltage signal from the device was pre-amplified with a low noise amplifier (SR560, Stanford Research Systems). For the details of the data processing, see S1 in SM.

**Calculation of magneto-chirality and toroidal moment.** The calculation of the magneto-chirality and the toroidal moment is performed using relevant data such as magnetization, modulation period of helical magnetic structure and lattice constants extracted from Refs. (*21*, *23*). See S4 in SM for the detailed descriptions.

**Acknowledgements**
We thank D. Nakamura and Y. Taguchi for fruitful discussions.

**Author contributions**
YT conceived and supervised the project. DY performed device fabrication, transport measurements, data analysis/visualization and calculation. AK grew single crystals. DY and YT wrote the original draft. NN reviewed and edited the draft. All the authors discussed the results and commented on the manuscript.

**References**

1.  Y. Tokura, S. Seki, N. Nagaosa, Multiferroics of spin origin. *Rep. Prog. Phys.* **77**, 076501 (2014).

2. H. Katsura, N. Nagaosa, A. V. Balatsky, Spin current and magnetoelectric effect in noncollinear magnets. *Phys. Rev. Lett.* **95**, 057205 (2005).

3. M. Mostovoy, Ferroelectricity in spiral magnets. *Phys. Rev. Lett.* **96**, 067601 (2006).

4. I. A. Sergienko, E. Dagotto, Role of the Dzyaloshinskii-Moriya interaction in multiferroic perovskites. *Phys. Rev. B* **73**, 094434 (2006).

5. Y. Yamasaki, H. Sagayama, T. Goto, M. Matsuura, K. Hirota, T. Arima, Y. Tokura, Electric control of spin helicity in a magnetic ferroelectric. *Phys. Rev. Lett.* **98**, 147204 (2007).

6. Y. Yamasaki, S. Myasaka, Y. Kaneko, J.-P. He, T. Arima, Y. Tokura, Magnetic reversal of the ferroelectric polarization in a multiferroic spinel oxide. *Phys. Rev. Lett.* **96**, 207204 (2006).

7. N. T. Bechler, J. Masell, Heritronics as a potential building block for classical and unconventional computing. *Neuromorph. Comput. Eng.* **3**, 034003 (2023).

8. N. Jiang, Y. Nii, H. Arisawa, E. Saitoh, Y. Onose, Electric current control of spin helicity in an itinerant helimagnet. *Nat. Commun.* **11**, 1601 (2020).

9. H. Masuda, T. Seki, J. Ohe, Y. Nii, H. Masuda, K. Takanashi, Y. Onose, Room temperature chirality switching and detection in a helimagnetic $MnAu_2$ thin film. *Nat. Commun.* **15**, 1999 (2024).

10. G. L. J. A. Rikken, J. Folling, P. Wyder, Electrical magnetochiral anisotropy. *Phys. Rev. Lett.* **87**, 236602 (2001).

11. F. Pop, P. Auban-Senzier, E. Canadell, G. L. J. A. Rikken, N. Avarvari, Electrical magnetochiral anisotropy in a bulk chiral molecular conductor. *Nat. Commun.* **5**, 3757 (2014).

12. D. Nakamura, M. K. Lee, K. Karube, M. Mochizuki, N. Nagaosa, Y. Tokura, Y. Taguchi, Nonreciprocal transport in a room-temperature chiral magnet. arXiv:2412.02272v1 (2024).

13. H. Ishizuka, N. Nagaosa, Anomalous electrical magnetochiral effect by chiral spin-cluster scattering. *Nat. Commun.* **11**, 2986 (2020).

14. Y. Tokura, Multiferroics as quantum electromagnets. *Science* **312**, 5779, 1481-1482 (2006).

15. N. A. Spaldin, M. Fiebig, M. Mostovoy, The toroidal moment in condensed-matter physics and its relation to the magnetoelectric effect. *J. Phys. Condens. Matter* **20**, 434203 (2008).

16. G. Venturini, D. Fruchart, B. Malaman, Incommensurate magnetic structures of $RMn_6Sn_6$ (R = Sc, Y, Lu) compounds from neutron diffraction study. *J. Alloys Compd.* **236**, 102-110 (1996).

17. A. Matsuo, K. Suga, K. Kindo, L. Zhang, E. Brück, K. H. J. Buschow, F.R. de Boer, C. Lefèvre, G. Venturini, Study of the Mn–Mn exchange interactions in single crystals of $RMn_6Sn_6$ compounds with R = Sc, Y and Lu. *J. Alloys Compd.* **408-412**, 110-113 (2006).

18. K. Uhlířová, V. Sechovský, F.R. de Boer, S. Yoshii, T. Yamamoto, M. Hagiwara, C. Lefèvre, G. Venturini, Magnetic properties and Hall effect of single-crystalline $YMn_6Sn_6$. *J. Magn. Magn. Mater.* **310**, 1747-1749 (2007).

19. A. A. Bykov, Y. O. Chetverikov, A. N. Pirogov, S. V. Grigor'ev, Quasi-two-dimensional character of the magnetic order-disorder transition in $YMn_6Sn_6$. *JETP Lett.* **101**, 699-702 (2015).

20. N. J. Ghimire, R. L. Dally, L. Poudel, D. C. Jones, D. Michel, N. Thapa Magar, M. Blemel, M. A. McGuire, J. S. Jiang, J. F. Mitchell, J. W. Lynn, I. I. Mazin, Competing magnetic phases and fluctuation-driven scalar spin chirality in the kagome metal $YMn_6Sn_6$. *Sci. Adv.* **6**, eabe2680 (2020).

21. Q. Wang, K. J. Neubauer, C. Duan, Q. Yin, S. Fujitsu, H. Hosono, F. Ye, R. Zhang, S. Chi, K. Krycka, H. Lei, P. Dai, Field-induced topological Hall effect and double-fan spin structure with a *c*-axis component in the metallic kagome antiferromagnetic compound $YMn_6Sn_6$. *Phys. Rev. B* **103**, 014416 (2021).

22. H. Zhang, X. Feng, T. Heitmann, A. I. Kolesnikov, M. B. Stone, Y.-M. Lu, X. Ke, Topological magnon bands in a room-temperature kagome magnet. *Phys. Rev. B* **101**, 100405(R) (2020).

23. R. L. Dally, J. W. Lynn, N. J. Ghimire, D. Michel, P. Siegfried, I. I. Mazin, Chiral properties of the zero-field spiral state and field-induced magnetic phases of the itinerant kagome metal $YMn_6Sn_6$. *Phys. Rev. B* **103**, 094413 (2021).

24. A. Kitaori, J. S. White, N. Kanazawa, V. Ukleev, D. Singh, Y. Furukawa, T.-h. Arima, N. Nagaosa, Y. Tokura, Doping control of magnetism and emergent electromagnetic induction in high-temperature helimagnets. *Phys. Rev. B* **107**, 024406 (2023).

25. Y. Tokura, N. Nagaosa, Nonreciprocal responces from noncentrosymmetric quantum materials. *Nat. Commun.* **9**, 3740 (2018).

26. P. Wadley, B. Howells, J. Železný, C. Andrews, V. Hills, R. P. Campion, V. Novák, K. Olejník, F. Maccherozzi, S. S. Dhesi, S. Y. Martin, T. Wagner, J. Wunderlich, F. Freimuth, Y. Mokrousov, J. Kuneš, J. S. Chauhan, M. J. Grzybowski, A. W. Rushforth, K. W. Edmonds, B. L. Gallagher, T. Jungwirth, Electrical switching of an antiferromagnet. *Science* **351**, 6273, 587-590 (2016).

27. H. Isobe, N. Nagaosa, Toroidal scattering and nonreciprocal transport by magnetic impurities. *J. Phys. Soc. Jpn.* **91**, 115001 (2022).

28. T. Yokouchi, N. Kanazawa, A. Kikkawa, D. Morikawa, K. Shibata, T. Arima, Y. Taguchi, F. Kagawa, Y. Tokura, Electrical magnetochiral effect induced by chiral spin fluctuations. *Nat. Commun.* **8**, 866 (2017).

29. S. Onoda, N. Nagaosa, Chiral spin pairing in helical magnets. Phys. Rev. Lett. **99**, 027206 (2007).

30. N. Nagaosa, Y. Tokura, Topological properties and dynamics of magnetic skyrmions. *Nat. Nanotechnol.* **8**, 899–911 (2013).

31. Y. Tokura, N. Kanazawa, Magnetic skyrmion materials. *Chem. Rev.* **121**, 2857-2897 (2021).

32. K. Shibata, X. Z. Yu, T. Hara, D. Morikawa, N. Kanazawa, K. Kimoto, S. Ishiwata, Y. Matsui, Y. Tokura, Towards control of the size and helicity of skyrmions in helimagnetic alloys by spin–orbit coupling. *Nat. Nanotechnol.* **8**, 723-728 (2013).

33. Y. Liu, B. Yang, X. Guo, S. Picozzi, Y. Yan, Modulation of skyrmion helicity by competition between Dzyaloshinskii-Moriya interaction and magnetic frustration. *Phys. Rev. B* **109**, 094431 (2024).

34. K. Momma, F. Izumi, VESTA 3 for three-dimentional visualization of crystal, volumetric and morphology data. *J. Appl. Cryst.* **44**, 1272-1276 (2011).

35. T. Ideue, K. Hamamoto, S. Koshikawa, M. Ezawa, S. Shimizu, Y. Kaneko, Y. Tokura, N. Nagaosa, Y. Iwasa, Bulk rectification effect in a polar semiconductor. *Nat. Phys.* **13**, 578-583 (2017).

36. R. Aoki, Y. Kousaka, Y. Togawa, Anomalous nonreciprocal electrical transport on chiral magnetic order. *Phys. Rev. Lett*. **122**, 057206 (2019).

37. C. Zhang, X. Yuan, J. Zhang, P. Leng, Y. Mou, Z. Ni, H. Zhang, C. Yu, Y. Yang, F. Xiu, Thermoelectric origin of giant nonreciprocal charge transport in NbAs nanobelts. *Phys. Rev. Appl.* **15**, 034084 (2021).

# Supplementally Materials for

# Magnetoelectric control of spin helicity and nonreciprocal charge transport in a multiferroic metal

Daiki Yamaguchi[1,2*], Aki Kitaori[1,3], Naoto Nagaosa[2,4], Yoshinori Tokura[1,2,5*]

[1]*Department of Applied Physics, The University of Tokyo, Tokyo 113-8656, Japan*
[2]*RIKEN Center for Emergent Matter Science (CEMS), Wako 351-0198, Japan*
[3]*Institute of Engineering Innovation, The University of Tokyo, Tokyo 113-0032, Japan*
[4]*RIKEN Fundamental Quantum Science Program, Wako 351-0198, Japan*
[5]*Tokyo College, The University of Tokyo, Tokyo 113-8656, Japan*

DY*: dai.20.yama.011-1@g.ecc.u-tokyo.ac.jp,
AK: kitaori@ap.t.u-tokyo.ac.jp,
NN: nagaosa@riken.jp,
YT*: tokura@riken.jp

* To whom correspondence should be addressed.

S1: Analysis procedure of nonreciprocal resistivity (NRR) data

The nonreciprocal charge transport in metals is understood as the current nonlinear term in the $I$-$V$ curve, where $I$ is the electric current and $V$ is the induced voltage drop. This $I$-$V$ profile is expressed as,

$$V = [R_0 + R_{\mathrm{ch}}(I)]I, \qquad (S1)$$

where $R_0$ is a constant and $R_{\mathrm{ch}}$ is $I$ dependent component of resistance. Therefore, the resistivity up to the first order of $j$ (current density) is,

$$\rho(j) = \rho_0 + \rho_{\mathrm{ch}} j, \qquad (S2)$$

where $\rho_0$ and $\rho_{\mathrm{ch}}$ are constants. Here, the nonreciprocal resistivity (NRR) is defined as an additional term ($\rho_{\mathrm{ch}} j$) in the resistivity. The NRR can be measured in the lock-in measurement, where ac current density ($j = I/S = j_{\mathrm{ac}} \sin(\omega t)$) with frequency ($f = \omega/2\pi$) is applied. $S$ is the cross section of the sample. Plugging the ac current into Eq. (S1) reads, up to the first order of $R_{\mathrm{ch}}(I) = R_1 I$,

$$V = R_0 I + R_1 I^2 = R_0 I_{\mathrm{ac}} \sin(\omega t) + R_1 I_{\mathrm{ac}}^2 \sin^2(\omega t) = R_0 I_{\mathrm{ac}} \sin(\omega t) + \frac{R_1 I_{\mathrm{ac}}^2}{2} - \frac{R_1 I_{\mathrm{ac}}^2}{2}\cos(2\omega t). \quad (S3)$$

Thus, using the electrode distance $l$, the resistivity is represented as $\rho_0 = R_0 S/l$ and $\rho_1 = R_1 S/2l = \rho_{\mathrm{ch}}$,

$$\rho(j_{\mathrm{ac}}) = \frac{V}{I_{\mathrm{ac}}} \cdot \frac{S}{l} = \rho_0 \sin(\omega t) + \rho_1 j_{\mathrm{ac}} - \rho_1 j_{\mathrm{ac}} \cos(2\omega t), \qquad (S4)$$

where the last term is the second harmonic resistivity. The imaginary (cosine) part of the second harmonic ($2f$) resistivity (Im $\rho_{zz}^{2f} = -\rho_{\mathrm{ch}} j_{\mathrm{ac}}$) is obtained as a raw signal in the lock-in measurement.

The raw signal Im $\rho_{zz}^{2f}$ includes several NRR components, which are not from the spin-helicity origin. In other words, irrelevant background NRR signals should be subtracted to focus on the NRR of the spin-helicity origin. As illustrated in Fig. S1, the raw signal of Im $\rho_{zz}^{2f}$ is finite in zero magnetic field, where NRR of spin-helicity origin is anticipated to be zero. The magnetic field dependence of clockwise (CW) and counterclockwise (CCW) overlap in the proper screw (PS) region and differ in the multiferroic transverse conical (TC) region. Therefore, we defined background as an average of CW and CCW signals. Thus, the CW and CCW curves are exactly symmetric, by definition. The background is found to have both field symmetric and asymmetric components. The symmetric backgrounds could be the effects of nonuniformity and electrode geometry (8, 35, 36). Meanwhile, the asymmetric backgrounds may be partly due to the Nernst effect from thermal gradient stemming from the different contact resistivity of the electrodes (*37*). This contribution of the Nernst effect is related to the device geometry. Figures S1G and S1H are the scanning electron microscopy (SEM) images of two different focused ion beam (FIB) fabricated devices. The widths are different between the two Hall-bar shaped devices. The background is more asymmetric in the wider device (Fig. S1G), which indicates that the effect of thermal gradient is suppressed in the narrow bar-shaped device (Fig. S1H). The background subtracted data of these two devices are similar regardless of the different field dependence of the backgrounds (Figs. S1 B and F). This indicates that the present procedure of the background subtraction is reasonable to focus on the NRR of spin-helicity origin.

The magnetoelectric (ME)-poling current density $j_{pol}$ is different between Figs. S1B and S1D. In the small $j_{pol}$, the spin-helicity control is not successful and the resulting NRR is zero. It is also confirmed from the overlapping signals of CW and CCW raw data in Fig. S1C.

S2: ac current density dependence of NRR

As anticipated from Eq. (S4), the NRR should be proportional to $j_{ac}$. Figure S2 confirms this $j_{ac}$ proportional nature of the NRR measured at 50 K with various $j_{ac}$ after the ME-poling with $j_{pol} = 1\times10^9$ Am$^{-2}$. The NRR is proportional to $j_{ac}$ both for the peak at around 2.3 T and the averaged NRR value in the TC phase.

S3: Magnetochiral (MCh)-poling and NRR sensing in longitudinal conical (LC) state

In the MCh-poling, the dc current and the magnetic field are both applied parallel to the *c*-axis (//*q*). Firstly, we applied the magnetic field along the *c*-axis up to 7 T at 270 K, and the high dc current density $j_{pol}$ along the *c*-axis. Here, $j_{pol}$ in the opposite direction is anticipated to realize the opposite spin helicity in the PS($B = 0$)/LC phase in the end. We define the spin helicity of the PS/LC phase as right-handed (RH) and left-handed (LH) after the MCh-poling with positive and negative $j_{pol}$, respectively (see Figs. 1 E and F in the main text). The magnetic field was slowly removed at a reduction rate of 1.2 mTs$^{-1}$ from 7 T to 5 T, while the current of $j_{pol}$ was kept flowing, enabling to traverse the phase boundary from the FF phase into the PS phase (Fig. S3A).

The spin helicity in the PS/LC phase was detected by the NRR arising from the electrical magnetochiral effect (eMChE). The NRR sensing was performed at 50 K between $\pm 9$ T (Fig. S3A). The background subtraction described in S1 is also applied in this case. The opposite-sign nature between the RH and LH curves in Fig. S3B indicates a successful operation of MCh-poling in YMn$_6$Sn$_6$. Note that in the case of eMChE, the sign change in magnetic field changes the spin helicity. Therefore, the sign of NRR is opposite between the positive and negative magnetic field regions for each (RH, LH) curve (Fig. S3B).

S4: Calculation of magneto-chirality and toroidal moment

The calculations of the magneto-chirality $\chi_v \cdot \boldsymbol{M} = (\boldsymbol{S}_i \times \boldsymbol{S}_j) \cdot \boldsymbol{M}$ and the toroidal moment $\boldsymbol{T} \sim \boldsymbol{P} \times \boldsymbol{M} \sim [\hat{\boldsymbol{q}} \times (\boldsymbol{S}_i \times \boldsymbol{S}_j)] \times \boldsymbol{M}$ along the $c$-axis ($//z//\boldsymbol{q}$) are performed under the assumption of the single helical pitch in the TC structure. The actual double helix structure is approximated by the single helix presentation and the interlayer Mn-Mn distance is approximated to be $c(T)/2$, which do not essentially change the discussion of the results described in the main text.

The Mn moment along the cone-axis of the TC structure at each magnetic field angle $\theta$ is estimated as,

$$M(\theta, T) = \sqrt{M(\theta = 90°, T)\cos^2\theta + M(\theta = 0°, T)\sin^2\theta}. \tag{S5}$$

The full moment of Mn $m(\theta, T = 2\text{ K})$ at each $\theta$ is also calculated in the same manner to consider the anisotropy of $m$. $M(\theta, T)$ is the projection of $m$ onto the cone-axis, and its magnitude depends on the strength of the external magnetic field $B$. Thus, the cone-opening angle $\phi$ can be calculated from $m$ and $M$,

$$\cos\phi(\theta, T) = \frac{M(\theta, T)}{m(\theta)}. \tag{S6}$$

From the $k_{z,1}$ and $k_{z,2}$ values (two spin modulation periods of double helix in reciprocal space), the cycloidal spin rotation angle is,

$$\varphi_j - \varphi_i = \frac{k_{z,1} + k_{z,2}}{2}\pi, \tag{S7}$$

assuming that the inter-layer Mn-Mn distance is $c(T)/2$ and that the spin rotation is homogeneous (single helix), for simplicity. The spin at site $i$ is expressed as (Fig. S4A),

$$\boldsymbol{S}_i = m \begin{pmatrix} \sin\phi\sin\varphi_i\sin\theta + \cos\phi\cos\theta \\ \sin\phi\cos\varphi_i \\ \sin\phi\sin\varphi_i\cos\theta - \cos\phi\sin\theta \end{pmatrix}. \tag{S8}$$

Therefore, the vector spin chirality $\chi_v^{ij} = \boldsymbol{S}_i \times \boldsymbol{S}_j$ is,

$$\boldsymbol{S}_i \times \boldsymbol{S}_j = m^2 \begin{pmatrix} \sin^2\phi \cos\theta \sin(\varphi_j - \varphi_i) + \sin\phi \cos\phi \sin\theta (\cos\varphi_j - \cos\varphi_i) \\ -\sin\phi \cos\phi (\sin\varphi_j - \sin\varphi_i) \\ -\sin^2\phi \sin\theta \sin(\varphi_j - \varphi_i) + \sin\phi \cos\phi \cos\theta (\cos\varphi_j - \cos\varphi_i) \end{pmatrix}. \quad (S9)$$

When we sum up along all $(i,j)$, the terms $(\sin\varphi_j - \sin\varphi_i)$ and $(\cos\varphi_j - \cos\varphi_i)$ become zero. Therefore, the macroscopically finite magneto-chirality $\chi_v \cdot \boldsymbol{M}$ is,

$$\chi_v \cdot \boldsymbol{M} = m^2 \sin^2\phi \begin{pmatrix} \cos\theta \sin(\varphi_j - \varphi_i) \\ 0 \\ -\sin\theta \sin(\varphi_j - \varphi_i) \end{pmatrix} \cdot M \begin{pmatrix} \cos\theta \\ 0 \\ \sin\theta \end{pmatrix} = m^2 M \sin^2\phi \sin(\varphi_j - \varphi_i). \quad (S10)$$

And the toroidal moment is,

$$\boldsymbol{T} \sim (\hat{\boldsymbol{q}} \times \chi_v) \times \boldsymbol{M} = \left[ \begin{pmatrix} 0 \\ 0 \\ \frac{c}{2} \end{pmatrix} \times m^2 \sin^2\phi \begin{pmatrix} \cos\theta \sin(\varphi_j - \varphi_i) \\ 0 \\ -\sin\theta \sin(\varphi_j - \varphi_i) \end{pmatrix} \right] \times M \begin{pmatrix} \cos\theta \\ 0 \\ \sin\theta \end{pmatrix}$$

$$= \frac{c}{2} m^2 M \sin^2\phi \cos\theta \sin(\varphi_j - \varphi_i) \begin{pmatrix} \sin\theta \\ 0 \\ -\cos\theta \end{pmatrix}.$$

$$(S11)$$

Thus, the component along the z-axis $T_z$ reads,

$$T_z \sim -\frac{c}{2} m^2 M \sin^2\phi \cos^2\theta \sin(\varphi_j - \varphi_i).$$

The relevant experimental data needed in the numerical calculation of $\chi_v \cdot \boldsymbol{M}$ and $T_z$ are taken from previous reports. $M(\theta = 0°, T)$, $M(\theta = 90°, T)$, $m(\theta = 0°, T = 2\text{ K})$ and $m(\theta = 90°, T = 2\text{ K})$ are taken from ref. 21. $k_{z,1}(T)$, $k_{z,2}(T)$ and $c(T)$ are taken from ref. 23. The resulting temperature dependences at each $\theta$ in 6 T are shown in Figs. S4 B and C. The same calculation is also performed for $B = 3$ T and 1.5 T ($15° \leqq \theta \leqq 90°$). The $\theta$ dependences are extracted from these results and used in the analysis in Fig. 3 in the main text. There, $\chi_v \cdot \boldsymbol{M}$ and $T_z$ are normalized by the values of $\chi_v \cdot \boldsymbol{M}(\theta = 0°, B = 6\text{ T})$ and $T_z(\theta = 0°, B = 6\text{ T})$, respectively.

**References**

35. T. Ideue, K. Hamamoto, S. Koshikawa, M. Ezawa, S. Shimizu, Y. Kaneko, Y. Tokura, N. Nagaosa, Y. Iwasa, Bulk rectification effect in a polar semiconductor. *Nat. Phys.* **13**, 578-583 (2017).

36. R. Aoki, Y. Kousaka, Y. Togawa, Anomalous nonreciprocal electrical transport on chiral magnetic order. *Phys. Rev. Lett.* **122**, 057206 (2019).

37. C. Zhang, X. Yuan, J. Zhang, P. Leng, Y. Mou, Z. Ni, H. Zhang, C. Yu, Y. Yang, F. Xiu, Thermoelectric origin of giant nonreciprocal charge transport in NbAs nanobelts. *Phys. Rev. Appl.* **15**, 034084 (2021).

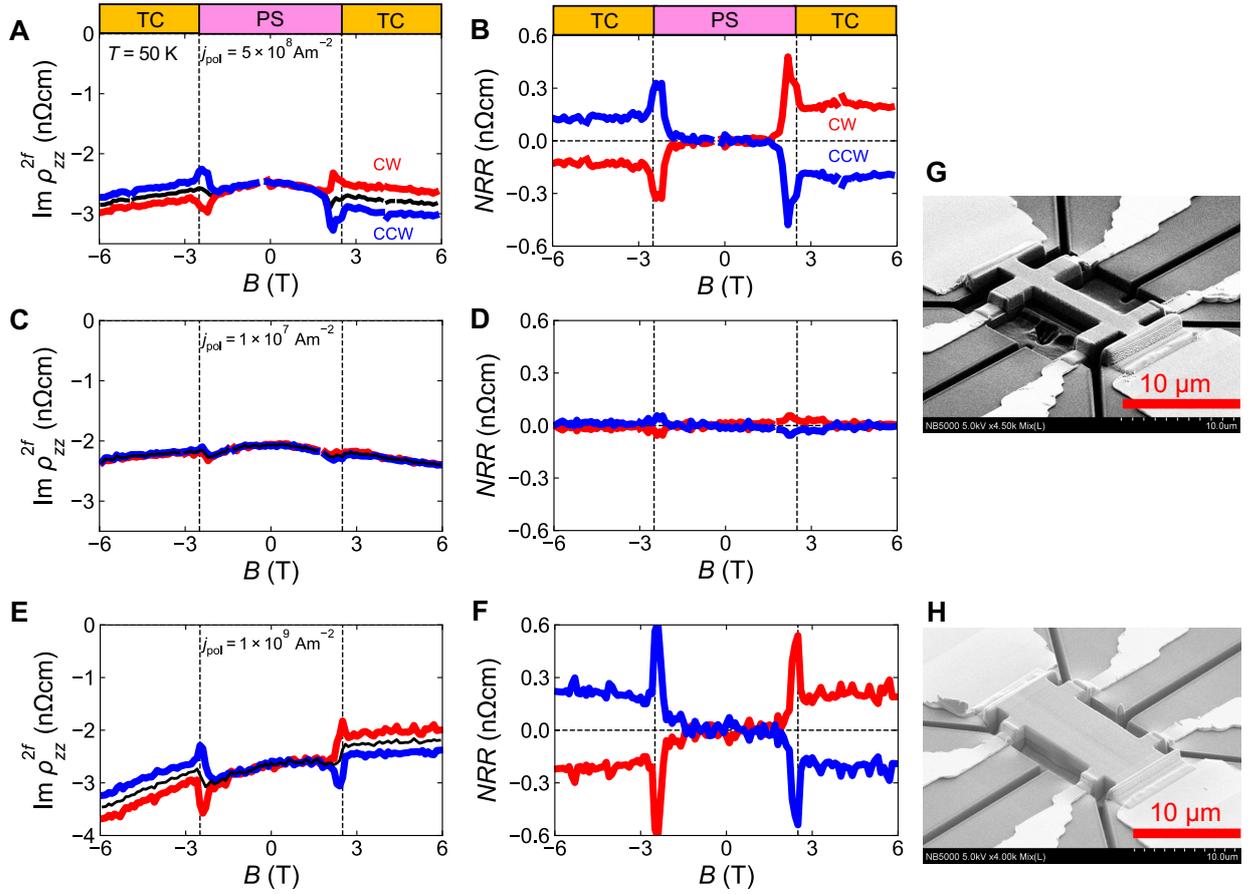

**Fig. S1. Raw and analyzed data of NRR after ME-poling in YMn$_6$Sn$_6$ FIB devices**

Magnetic field dependence of the raw signal of the second harmonic resistivity measured at 50 K with $j_{ac}$ = 5×10$^8$ Am$^{-2}$ after magnetoelectric (ME)-poling with (**A**) $j_{pol}$ = 5×10$^8$ Am$^{-2}$, (**C**) $j_{pol}$ = 1×10$^7$ Am$^{-2}$ and (**E**) $j_{pol}$ = 1×10$^9$ Am$^{-2}$. Red and blue lines correspond to positive (CW) and negative (CCW) $j_{pol}$, respectively. Black lines are the average of CW and CCW curves in each panel, which are defined as background curves. (**B**), (**D**) and (**F**) are background subtracted data of NRR for (A), (C) and (E), respectively. (**G**) Scanning electron microscopy (SEM) image of a focused ion beam (FIB) fabricated device, in which NRR in (A-D) are measured. (**H**) SEM image of a FIB device, in which NRR in (E, F) are measured.

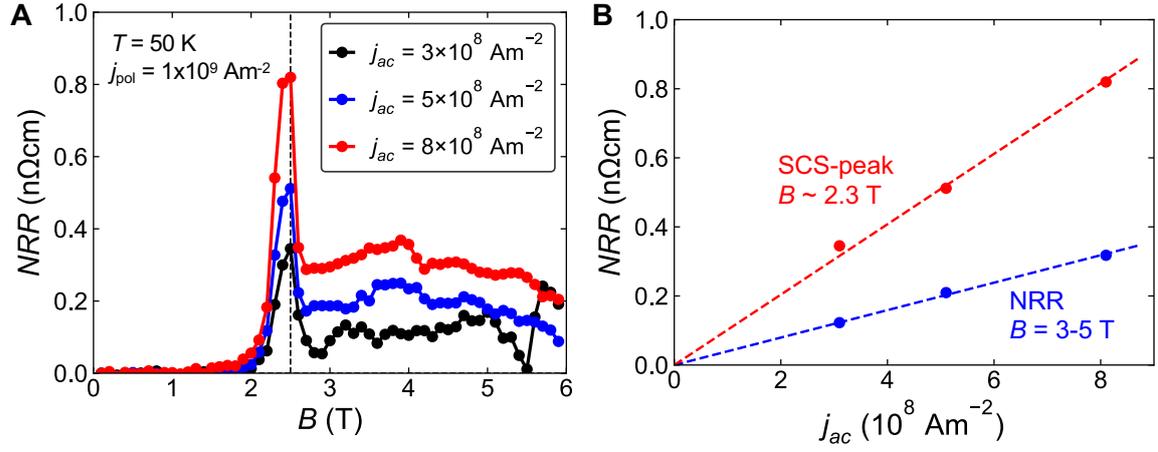

**Fig. S2. ac current density dependence of NRR**

(**A**) Magnetic field dependence of NRR measured at 50 K with various $j_{ac}$ after ME-poling with $j_{pol} = 1\times10^9$ Am$^{-2}$. (**B**) $j_{ac}$ dependence of NRR value at peak at PS-TC phase boundary (red) and value in TC phase (blue). The red and blue dashed lines are the linear fitting of NRR value and value in TC phase, respectively.

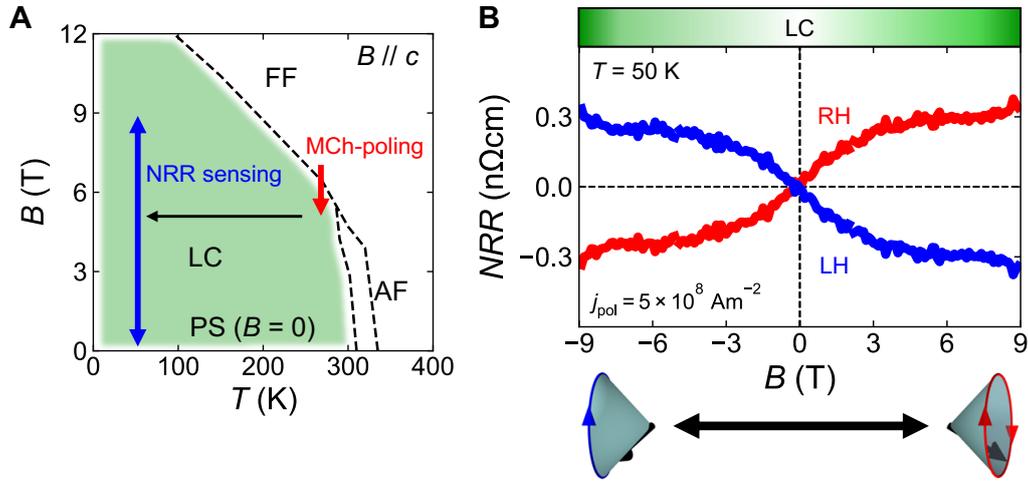

**Fig. S3. Magnetochiral (MCh)-poling and NRR sensing in $B//j//c$**

(**A**) MCh-poling (red arrow) and NRR sensing (blue arrow) procedures are indicated in the magnetic phase diagram. (**B**) Magnetic field dependence of the NRR measured at 50 K with $j_{ac}$ = **5×10$^8$ Am$^{-2}$** after MCh-poling with $j_{pol}$ = **+5×10$^8$ Am$^{-2}$** (red, RH) and $j_{pol}$ = **-5×10$^8$ Am$^{-2}$** (blue, LH). The schematics show the spin-helicity reversal between positive and negative magnetic field regions (the case of starting the field sweep from RH, 9 T).

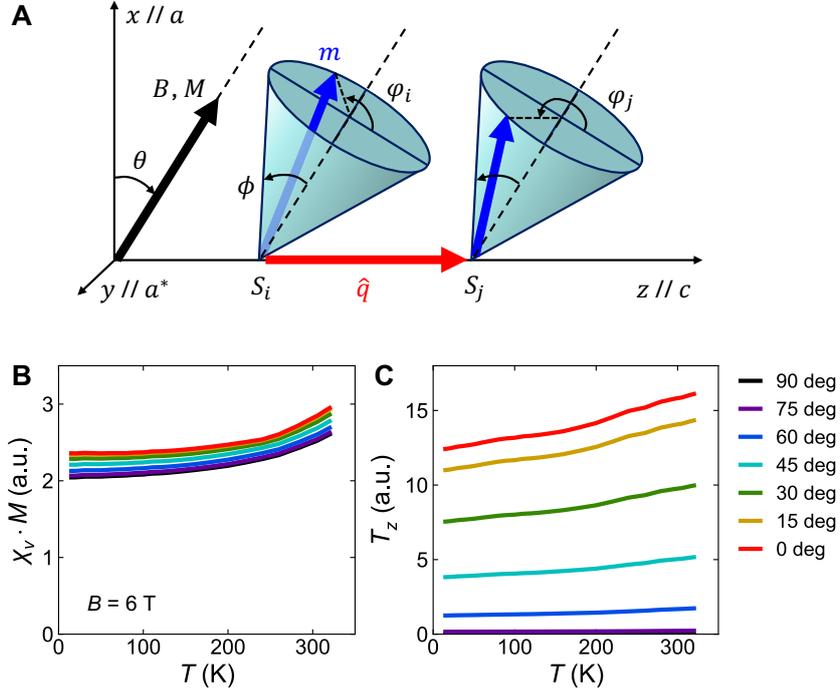

**Fig. S4. Calculation of magneto-chirality and toroidal moment**

(**A**) Schematic of the nearest-neighbor spins ($S_i$ and $S_j$) in conical order along the *c*-axis under the magnetic field direction apart from the *a*-axis by $\theta$. $m$, $\hat{q}$, $\phi$ and $\varphi_{i,j}$ are the full moment of Mn, the unit modulation wave vector connecting $S_i$ and $S_j$, the cone-opening angle measured from the cone-axis (dashed lines), and the rotation angles of the cycloidal component at sites *i* and *j*, respectively. Calculated temperature dependence of (**B**) magneto-chirality $\chi_v \cdot M$ and (**C**) toroidal moment along *z*-axis $T_z$ at each $\theta$.